\pdfoutput=1
\documentclass[amssymb,aps,twocolumn,showpacs,nofootinbib]{revtex4-1}
\usepackage[]{graphicx,subfig,float}
\usepackage[]{amsmath}
\usepackage{hyperref}
\usepackage{color}

\def\ket#1{| #1 \rangle}

\def\kb#1#2{| #1 \rangle\!\langle #2 |}
\def\II{1\!\mathrm{l}}

\def\cE{\mathcal{E}}
\def\cF{\mathcal{F}}
\def\cG{\mathcal{G}}

\def\cJ{\mathcal{J}}

\def\cN{\mathcal{N}}
\def\cO{\mathcal{O}}
\def\cP{\mathcal{P}}

\def\cS{\mathcal{S}}
\def\cT{\mathcal{T}}

\def\Tr{\mathrm{Tr}}

\def\eq#1{Eq.~\eqref{eq:#1}}
\def\fig#1{Fig.~\ref{fig:#1}}
\def\sec#1{Sec.~\ref{sec:#1}}

\begin{document}

\title{A Small Quantum Computer is Needed to Optimize Fault-Tolerant Protocols}
\author{Pavithran S. Iyer}
\email{Pavithran.Iyer.Sridharan@USherbrooke.ca}
\author{David Poulin}
\email{David.Poulin@USherbrooke.ca}
\affiliation{D\'epartement de Physique \& Institut Quantique, Universit\'e de Sherbrooke, Qu\'ebec, Canada}

\date{\today}

\begin{abstract}
As far as we know, a useful quantum computer will require fault-tolerant gates, and existing schemes demand a prohibitively large space and time overhead. We argue that a first generation quantum computer will be very valuable to design, test, and optimize fault-tolerant protocols  tailored to the noise processes of the hardware. Our argument is essentially a critical analysis of the current methods envisioned to optimize fault-tolerant schemes, which rely on hardware characterization, noise modelling, and numerical simulations. We show that, even within a very restricted set of noise models, error correction protocols depend strongly on the details of the noise model. Combined to the intrinsic difficulty of hardware characterization and of numerical simulations of fault-tolerant protocols, we arrive at the conclusion that the currently envisioned optimization cycle is of very limited scope. On the other hand, the direct characterization of a fault-tolerant scheme on a small quantum computer bypasses these difficulties, and could provide a bootstrapping path to full-scale fault-tolerant quantum computation.
\end{abstract}

\maketitle

\section{Motivation}

While we know that a quantum computer can in principle solve certain problems exponentially faster than the best known classical algorithms, a very large quantum computer is likely to be required to beat classical computers on a problem of intrinsic interest (as opposed to a made-up problem conceived to demonstrate a quantum advantage, e.g., \cite{A11,BBB16}). There are basically two reasons for this. First, classical computers are extremely large and fast. The world's fastest supercomputers operate at nearly 100 quadrillion (i.e. $10^{17}$) floating-point operations per second on a memory of nearly a quadrillion bytes. While this is largely achieved by parallelization, even the CPU used to write this article performs a few billion operations per second on a memory of a few tens of billions of bytes. In contrast, the typical clock rate of solid-state quantum computers enables a few million operations per second, and in this collection of articles we imagine an early generation of devices containing on the order of a thousand qubits. 

While these quantum clock rates and memories sizes may appear reasonably large, we must not forget that quantum systems are highly susceptible to noise, which bring us to the second reason. As far as we know, quantum algorithms need to be implemented fault-tolerantly to provide reliable answers. As a consequence, each logical qubit of the algorithm must be encoded in some quantum error-correcting code using  several (hundreds of) physical qubits, and each elementary gate in the quantum algorithm is implemented using several (thousands of) elementary gates on the physical hardware \cite{FMMM12}. Thus, the  noisy physical device described in the previous paragraph might at best produce a reliable quantum computer performing a thousand operations per second on a dozen qubits. 

One important research area in quantum information science is aimed at lowering this fault-tolerance overheads, i.e. finding better codes and fault-tolerant protocols which require fewer qubits, fewer gates, and achieve a better error suppression. While early studies in this area focused on ``featureless'' depolarizing noise, it has become clear that substantial gains can be achieved by taking into account specific details of the hardware in the protocol design \cite{ABD09, PJ08, WBP15, TBF17}. At the moment, this is done at a rather coarse level: the foremost example is biased noise models, where it is assumed that errors corresponding to Pauli $X$  matrices (bit flip) are much less likely than those corresponding to Pauli $Z$ matrices (phase flip). This biased noise model is motivated by qubits built from non-degenerate energy levels where a bit flip requires an energy exchange with the bath, so it is typically much slower than phase flips, which only require an entropy exchange. 

While the noise bias is one of many features which can colour a noise model, fault-tolerant protocols can be tailored to various other features. This research program thus naturally suggest an optimization cycle which combines 
\begin{enumerate}
\item Experimental noise characterization of device.
\item Noise modeling.
\item Fault tolerant protocol design tailored to model.
\item Numerical benchmark of protocol.
\end{enumerate} 
The main message of this article is that the above optimization cycle is not viable, and that given access to  a small quantum information processor, steps 1 and 4 could be combined into a single step: {\em Experimental benchmark of fault-tolerant protocol}. We are lead to this conclusion by three observations. First, experiments can only extract coarse information about the noise affecting the hardware. Second, the response of a fault-tolerant scheme depends strongly on the fine parameters of the noise. Third, the response of a fault-tolerant scheme to even simple noise models is computationally hard to predict. We will now elaborate on each of these observations.

A noise bias is but one of many features that a noise model can have. At the level of a single qubit, the evolution operator is described by 12 real parameters, the bias being only one of them. That number grows exponentially with the number of qubits, the growth being mainly attributed to the number of inequivalent ways in which errors can be correlated across different qubits. Temporal correlations and non-Markovian effects will further increase that number of parameters, resulting in an extremely high-dimensional noise model manifold. 

Thus, it is technically impossible to fully characterize the noise affecting more than, say, 3 qubits \cite{FIP97,KGK17}. Techniques have been developed over the past decade to extract coarse information about the noise inflicting a system \cite{WT14,MGE11,AJJ16,BWCO11,CMBL16}. The simplest of these techniques will describe the noise by a single parameter $0<p<1$, which gives some indication of its strength, and more elaborate schemes will provide more parameters \cite{KLDF16}. These parameters define hypersurfaces in the high-dimensional noise manifold, leaving many noise parameters unspecified. One is left to wonder if knowledge about these few parameters can be of any help in designing tailored fault-tolerant protocols. 

One of the key messages of this article is that unfortunately, no, there appears to be very little to be gain from such coarse information. This does not conflict with what we wrote above, about how knowledge of the noise bias has led to improved tailored protocols. In those examples, the hidden assumption was that the noise is biased but otherwise featureless. There exist other biased noise models exhibiting other types of correlations for which the tailored protocols fail. In other words, fixing some noise hypersurface while letting the other parameters fluctuate will result in vastly different noise models that react wildly differently to fault-tolerant protocols. To support these claims, we will  present in \sec{criticalparameters} numerical simulation results showing how the response of a given error correcting scheme can wildly fluctuate for noise models of equal strength. 

These results lead us to ask what are the critical parameters which most strongly affect the response of a fault-tolerant scheme. To investigate this question, we have used machine learning techniques to attempt to correlate the response of a fault-tolerance scheme to the parameters of the noise model. Our results will be presented in \sec{ML}. We have tried a few different machine learning algorithms and the critical parameters we found were more informative than generic noise strength measures, such as average infidelity or the diamond norm. Despite these relative improvements the accuracy of the predictions from machine learning algorithms remain poor. This provides further evidence that fine details of the noise model must be known to predict -- and eventually optimize -- the response of a fault tolerant scheme.

In \sec{numerics}, we will discuss the numerical difficulty of simulating an error correction process. While several problems related to classical and quantum error correction -- such as optimal decoding -- are notoriously hard computational problems \cite{BMT78,IP15}, the characterization of quantum protocols poses an extra computational challenge with no classical counterpart. This difficulty  stems from the computational hardness of simulating quantum mechanics. From that perspective, it is rather surprising that numerical simulations can be of any use to simulate large quantum error-correcting schemes, but the Gottesman-Knill theorem \cite{G98} provides a means to efficiently simulate simple noise models. However,  the assumptions of the Gottesman-Knill theorem pose severe limitations on the noise models which can be efficiently simulated, thus rendering numerical simulations rather useless for the design of  fault tolerant schemes tailored to physical noise models. 

In addition to this quantum hardness, the numerical characterization of error correcting schemes is plagued by the inherent difficulty of characterizing rare events. Indeed, the interest of a fault tolerant scheme is that it results in a very low logical fault-rate. Thus, understanding and characterizing such faults requires an extremely large number of simulations. In classical error correction, it is possible to use importance sampling methods which  enhance the probability of these rare events, see, e.g., \cite{RT09, CL15, B12}. Here again, quantum error correction poses an new challenge because quantum errors are generically not described by stochastic processes, and hence importance sampling methods do not directly apply. In \sec{IS}, we will present our attempts at developing importance sampling methods tailored to quantum processes. While we obtain some improvements over direct simulations, the number of simulations required for practical quantum computing applications remains prohibitively large.   


\section{Numerical simulation}
\label{sec:NS}

The evolution of a single qubit state $\rho$ over some fixed period of time can be described by a completely positive, trace preserving (CPTP) map $\cE(\rho) = \sum_{kk'} \chi_{kk'} \sigma_k \rho \sigma_k$ where $\sigma_k$ are the Pauli matrices and the complex $4\times 4$  matrix $\chi$ has unit trace \cite{WBC15}.
Furthermore, $\cE$ has 12 independent real parameters \cite{RSW02}. It can be shown that $\cE$ can always be obtained by considering a unitary evolution $U$ involving the qubit in state $\rho$ together with two additional ``environmental'' qubits initially in state $\ket 0$, i.e., $\cE(\rho) = \Tr_E\{U \rho\otimes \kb 00 U^\dagger\}$. 

We are interested in studying a wide range of physical noise models, so we choose to generate random single-qubit CPTP maps $\cE$. Note that there is no natural notion of uniform distribution over the space of CPTP maps\footnote{We choose a distribution which is unitarily invariant, but this leaves several parameters of the distribution unspecified.}. We can generate random single-qubit noise models $\cE$ using the equivalence to three-qubit unitary matrices $U$ described above. Specifically in this study, we  generate a three qubit Hamiltonian $H$ with gaussian distributed unit-variance entries and construct the unitary matrix $U=e^{i\delta H}$ where $\delta$ is a real parameters providing us with some handle on the noise ``strength''. 

To characterize the response of a fault-tolerant scheme to a given noise model $\cE$, we perform numerical simulations of the concatenated 7-qubit Steane code \cite{S96}. The Steane code encodes a single logical qubit and has minimal distance $d=3$, i.e., it can correct an arbitrary error on $t=\frac{d-1}2=1$ qubit. In a concatenated code \cite{F65}, we encode each of the 7 qubits making up the code in separate error correcting codes, resulting in a code which encodes a single logical qubit in $49=7^2$ physical qubits and with minimal distance $9=3^2$. The procedure can be repeated and we have simulated up to 4 levels of concatenation.

To simplify our task, we have assumed that the device only suffers from initial memory error, i.e., we assume that the gates used to error-correct are noiseless. While this is not a realistic assumption, it very significantly reduces the dimension of the noise model manifold. Indeed, a complete noise model would not only need to specified the single qubit CPTP map $\cE$ describing the noise suffered by an idle qubit, but would further specify a noise model of each unitary gate, measurement process, and state preparation. Thus, we can anticipate that  understanding  critical parameters in a complete noise model will be much more challenging than in the simplified model we adopt here, so our conclusions  remain perfectly valid despite this simplification.


Here we outline the steps in our numerical simulation,  technical details can be found in the appendix. We initialize the simulation in an 8-qubit maximally entangled state, $\rho_0$, between an encoded qubit in the Steane code and a reference qubit. While the reference qubit is noiseless, a single qubit channel $\cE_{0}$ is applied on each of the physical qubits of the Steane code, thus making up an i.i.d channel $\cE^{\otimes 7}_0$. The subscript 0 makes reference to the fact that these are physical noise models. In a suitable representation (see the appendix), this  7-qubit channel is a $4^7\times 4^7$ real matrix in tensor product form. The action of the noise on $\rho_{0}$ produces a 8-qubit state $\rho_{\rm noisy}$. We then apply the error-correction circuit to $\rho_{\rm noisy}$. An error correction circuit comprises of syndrome measurements corresponding to the six stabilizer generators $S_j$. We numerically compute the probability $\Pr(\pm) = \frac 12(1\pm \Tr\{S_j\rho_{\rm noisy}\})$ for each measurement outcome and choose the outcome at random following that distribution, resulting in a post-measurement state $\rho_{\rm noisy}^{s}$, which depends on the measured syndrome $s$.

Given a syndrome $s$, we choose the Pauli operator $Q$ that maximizes the fidelity $F(\rho_0,Q \rho_{\rm noisy}^s Q)$ to the initial noiseless state $\rho_{0}$. While there a priori appears to be $4^7$ Pauli matrices $Q$ to choose from, there are really only 4 distinct ones to choose from, corresponding to the 4 logical Pauli operators. The error-corrected state $Q_{\rm max} \rho_{\rm noisy}^s Q_{\rm max}$ encodes a noisy entangled state between the encoded qubit and the reference qubit. Thanks to the Jamio\l kowski-Choi isomorphism \cite{J72,C75}, knowledge of this state is equivalent to knowledge of the single-logical-qubit channel $\cE^s_1$ which has been applied to the logical qubit, conditioned on the syndrome $s$ which was observed. The simulation also yields the probability $\Pr(s)$ of the chosen syndrome. This terminates one instance of the simulation.

To simulate concatenation, we repeat the above procedure 7 times, yielding 7 single-qubit logical channels $\cE_1^{s_j}$. We then repeat the above procedure one last time using the noise model $\bigotimes_{j=1}^7 \cE_1^{s_j}$, which describes the noise model seen by the second concatenation layer, conditioned on the syndromes of the first layer.  This simulation results in a single-qubit logical channel $\cE^s_2$, where $s$ now denotes the collection of all level 1 syndrome as well as the level 2 syndrome: it comprises $7\times 6 + 6 = 48$ syndrome bits. Thus, we will sometimes refer to $s$ as the syndrome history.

After $\ell$ levels of concatenation, the average channel experienced by a logical qubit is $\overline \cE_{\ell} = \sum_s \Pr(s) \cE^s_{\ell}$. The range of this sum grows exponentially with $\ell$, so even for three levels of concatenation we are forced to sample the distribution $N$ times instead of evaluating the sum directly, which provides an estimate $\widetilde \cE_{\ell} = \frac{1}{N} \sum_j \cE^{s_j}_{\ell}$ of $\overline \cE_{\ell}$, where $s_j$ denotes the sampled syndromes. 

\section{Critical noise parameters}
\label{sec:criticalparameters}

A very coarse description of a noise model $\cE$ would be a single number specifying its ``noise level", or ``strength'', with strength 0 corresponding to a noiseless channel (the identity map). There are several inequivalent measures which are used to describe the strength. Some, such as the average infidelity, are efficiently accessible experimentally \cite{EAZ05,KLR08,WT14}. Others, like the diamond distance \cite{Kit97,W09,G05}, are more convenient mathematically but much more challenging to probe experimentally. Let us denote by $\cN$ such generic noise measure, i.e., $\cN(\cE)$ is the noise strength of the CPTP map $\cE$. 

Heuristically, the fault-tolerant accuracy threshold theorem \cite{AB08, KLZ96, K97, Pres98} states that, provided the noise strength $\cN_0 \equiv \cN(\cE_0)$ of the physical channel is less than a certain threshold value, the average logical noise strength $\overline \cN_\ell$ will decrease doubly exponentially with the level of concatenation $\ell$. This theorem is proved either assuming a stochastic noise model \cite{AGP07} -- in which case all metrics are essentially equivalent -- or using the diamond norm distance \cite{AB08,SDT07,AP09} in place of $\cN$. The theorem makes conservative assumptions about the nature of the noise model, and it at best provides very loose upper bounds on $\cN(\overline \cE_\ell)$. Upper bounds are of very little use in order to optimize a fault tolerant scheme, so we would like to develop a better understanding of  the behaviour of $\overline \cN_\ell$. 

In the previous section, we described a numerical procedure to sample logical channels $\cE^s_\ell$ corresponding to $\ell$ levels of concatenation. This gives us a mean to estimate, within statistical errors, the average logical noise strength. Note that there are two natural definitions of the average logical noise strength, either as the noise strength of the average channel \cite{CGFF17}
\begin{equation}
\cN(\overline \cE_\ell) = \cN\left(\sum_s \Pr(s) \cE_\ell^s\right),
\label{eq:N1}
\end{equation}
or as the average of the noise strength over the different syndromes \cite{GSCL16}
\begin{equation}
\overline{\cN( \cE_\ell)} = \sum_s \Pr(s) \cN(\cE_\ell^s).
\label{eq:N2}
\end{equation}
We have used both definitions in our numerical simulations, and this choice has quantitative but no qualitative effect on our conclusions. The results presented in the rest of this article use the measure of \eq{N2}, but we will continue to use the generic notation $\overline \cN_\ell$. As above, we will denote $\widetilde \cN_\ell$ the empirical estimate of $\overline \cN_\ell$.

\subsection{Standard error metrics}
\label{sec:StandardMetrics}

\begin{figure*}
\subfloat[]{\includegraphics[scale=0.14]{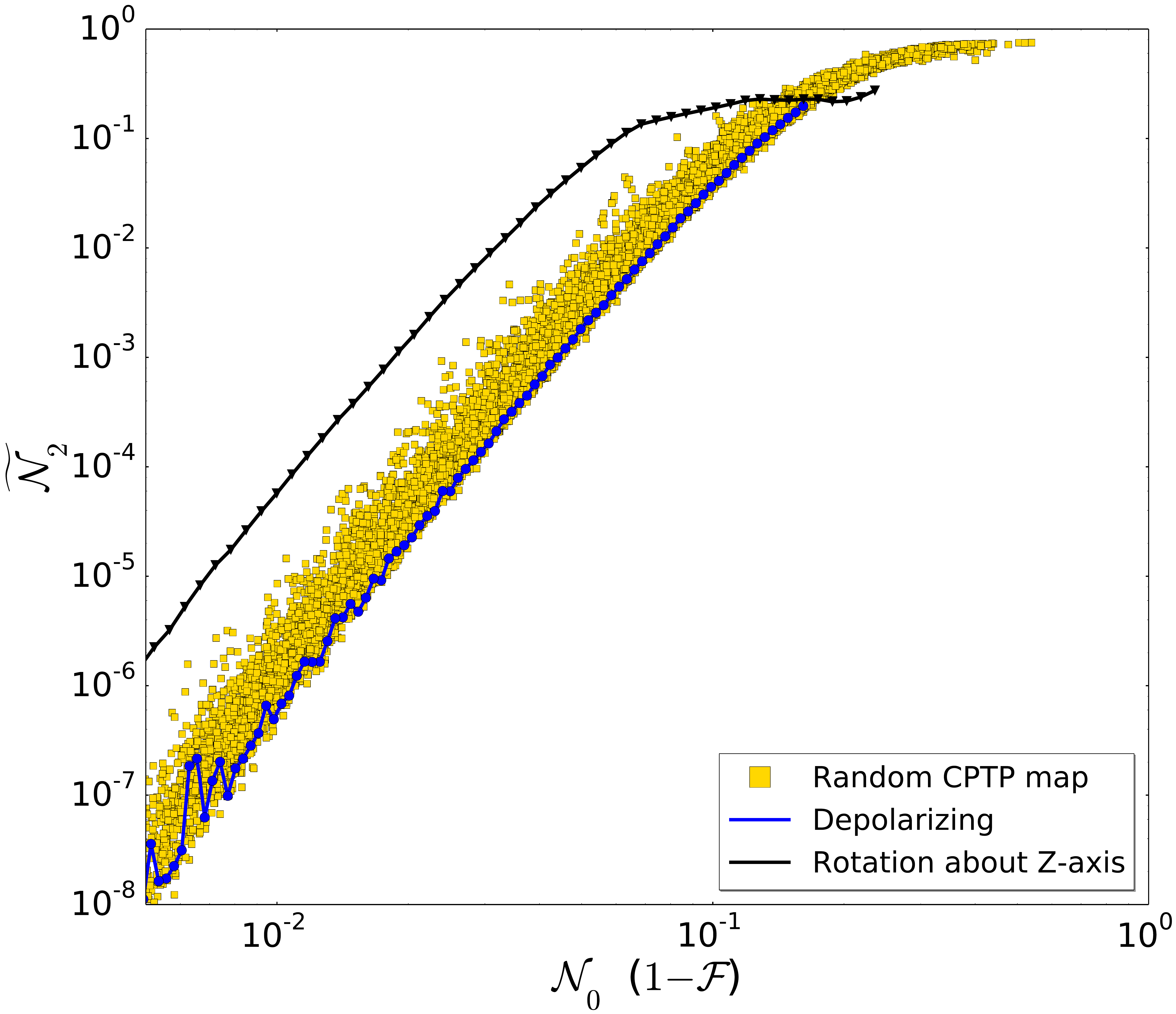}}
\subfloat[]{\includegraphics[scale=0.14]{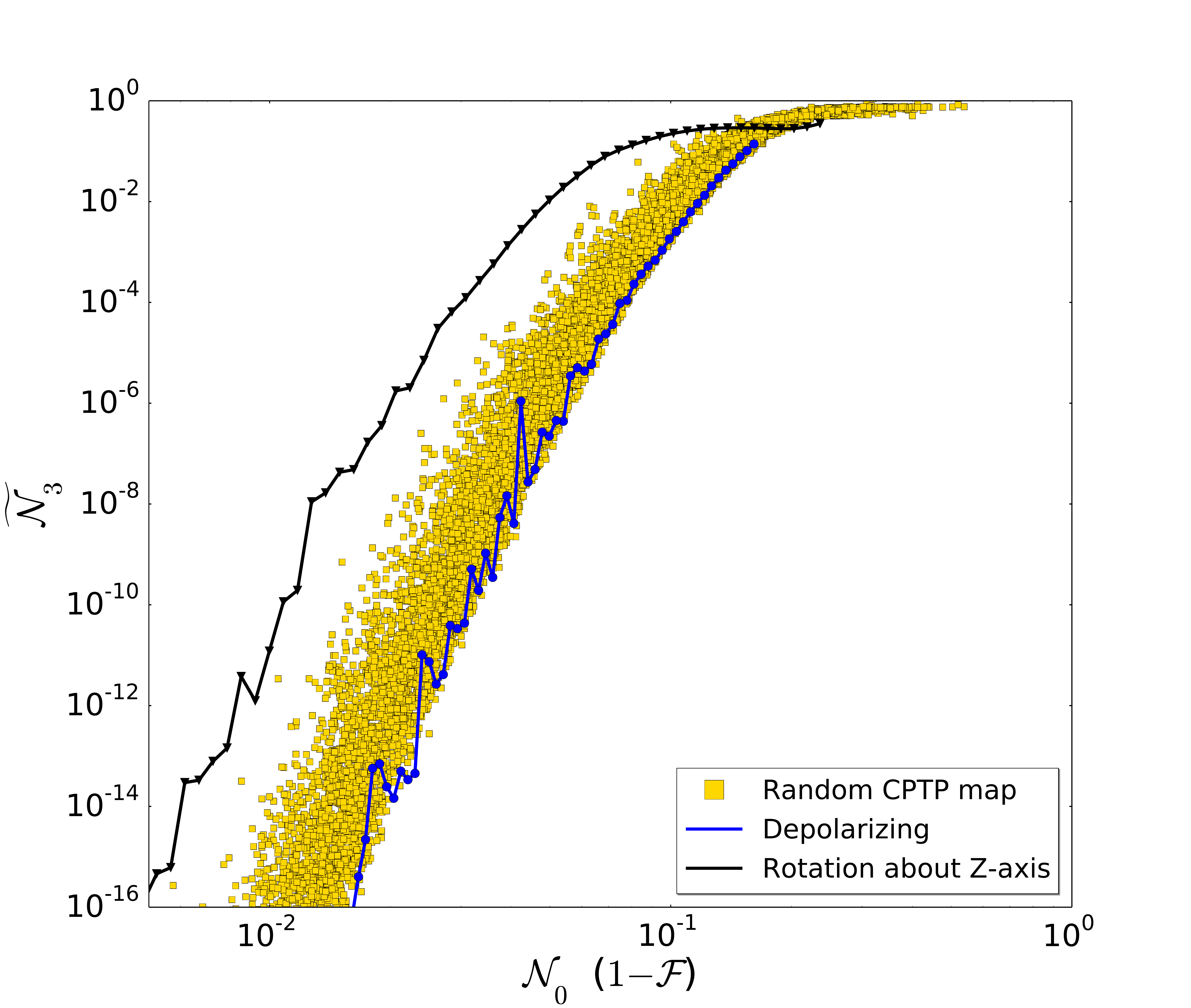}}

\subfloat[]{\includegraphics[scale=0.14]{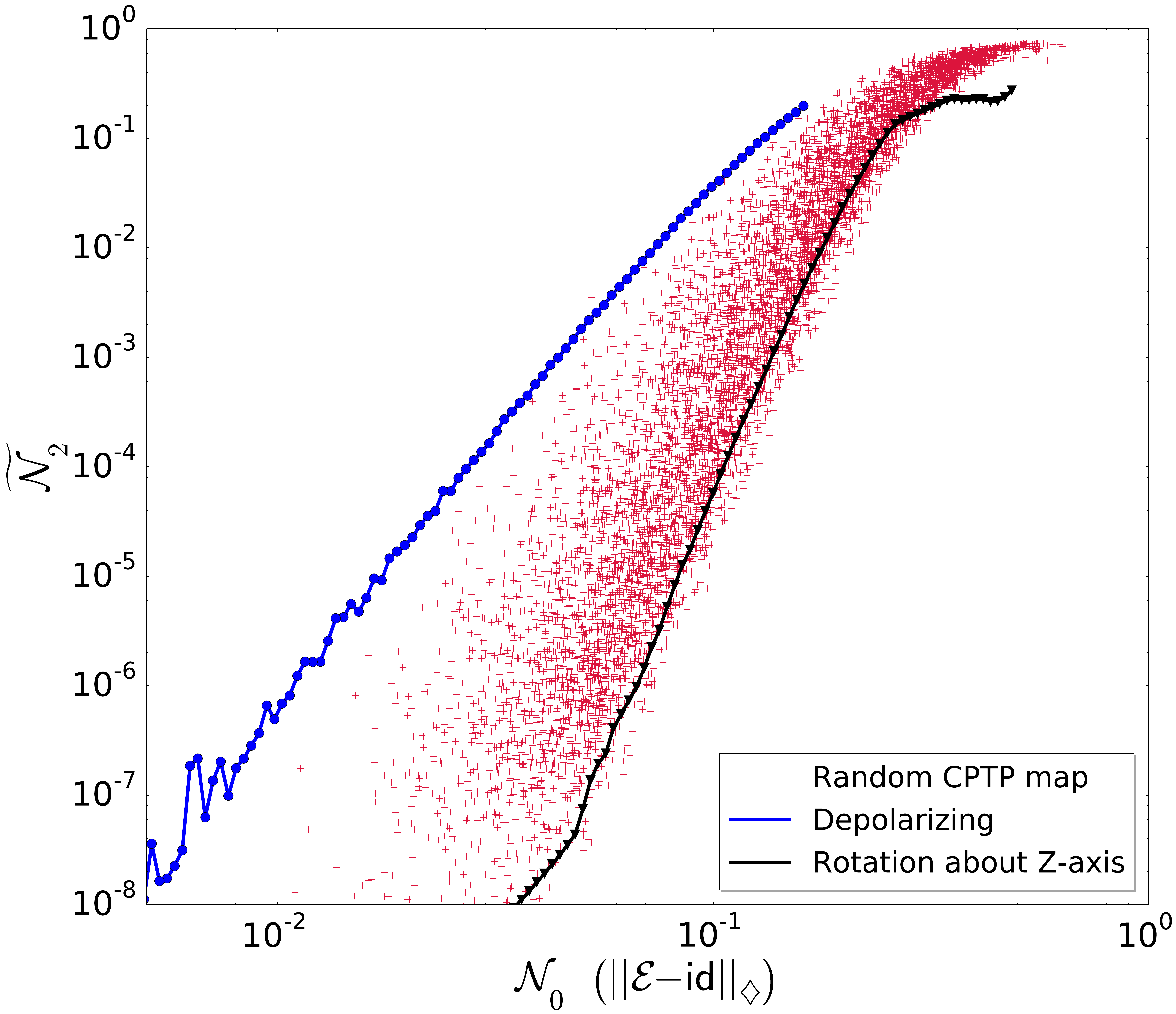}}
\subfloat[]{\includegraphics[scale=0.14]{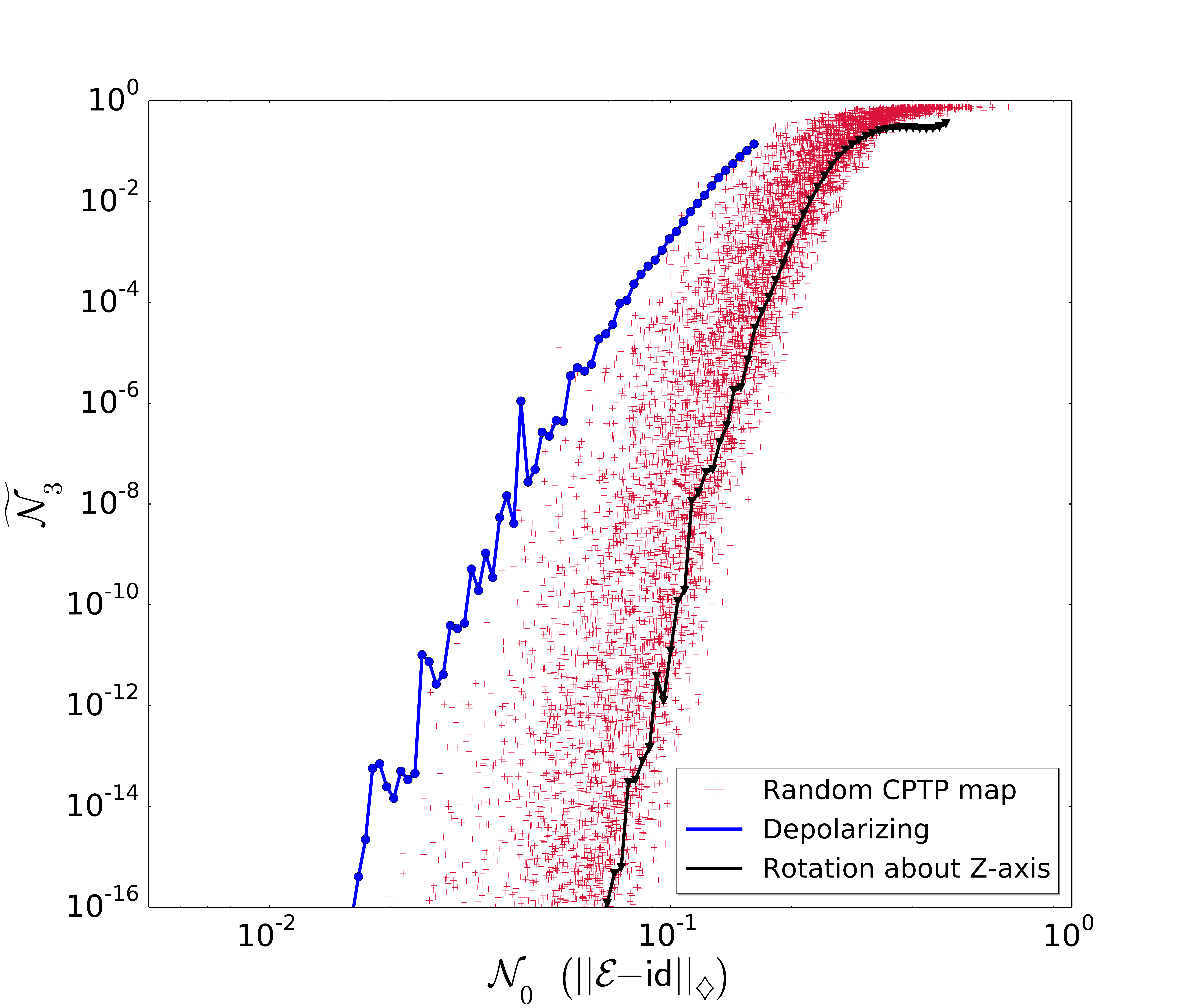}}
\caption{Average logical noise strength as a function of the physical noise strength. Each of the $12\times10^{4}$ dots correspond to a random channel and has been sampled $10^{4}$ times. Blue line corresponds to the depolarizing channel while the black line corresponds to the rotation channel. The logical noise strength $\widetilde \cN_\ell$ is measured using infidelity. Physical noise strength $\cN_0$ is measured using infidelity in a) and b), while it uses the diamond norm distance for c) and d). The number of concatenation levels is $\ell=2$ for a) and c) and $\ell=3$ for b) and d). The plots have a large scatter -- e.g., logical error rates vary by 10 orders of magnitude across channels with $\cN_{0} \sim 0.1$ in d) -- indicating that it is not possible to even crudely predict the average logical noise strength given only the physical noise strength.}
\label{fig:scatter}
\end{figure*}

Figure \ref{fig:scatter} shows the average logical noise strength as a function of the physical noise strength for a wide range of channels and using different measures of noise strength. What we observe is that the logical noise strength varies wildly for a fixed physical noise strength, which implies that estimating the logical noise strength given only the physical noise strength is doomed to yield extremely inaccurate estimates\footnote{There is a visible gap in the scatter plots, for instance the depolarizing channel is rather isolated in \fig{scatter} c). This is an artefact of the method we adapted to sample random channels, and as such does not reveal anything particularly deep. We have indeed used other sampling methods and found that this void disappears.}. We have used several combinations of noise measures -- infidelity, diamond norm distance, 1-norm distance, 2-norm distance, entropy, and worst case error --  (defined in Sec. \ref{app:naturalMetrics}) which all produced similar looking scatter plots. Infidelity was the best metric we found in terms of its ability to predict the behavior of the logical channel, but not by a significant margin.

Focusing on the graphs of \fig{scatter} c) and d), we reach the conclusion that depolarizing is amongst the worst noise model in the sense that most channels of equal strength result in much less logical noise. This is appealing since the vast majority of numerical simulations to date use the depolarizing channel and furthermore, many of the fault tolerance proofs use the depolarizing channel along with the diamond norm, so from this point of view these studies would provide a worst case scenario. However, using infidelity as our measure of noise strength as in \fig{scatter} a) and b) yields the opposite conclusion: the depolarizing channel is now amongst the best physical channels. This stresses the importance of  choosing an appropriate measure to report the accuracy of an experiment, and more generally motivates the search of critical parameters which best correlate with the logical noise strength. 

\subsection{Machine learning of critical parameters}
\label{sec:ML}

Figure \ref{fig:scatter} shows how the simple knowledge of the physical noise strength -- as measured by any of the standard metrics -- provides very little information about the response of a fault-tolerant scheme to a given noise model. This motivates the search for other critical parameters of the noise models, whose value enables us to better predict the behaviour of the induced logical noise. In this section, we will present our attempt at using machine learning techniques to find such critical parameters. The basic idea is to find a ``simple'' function of the channel parameters $f(\cE_0)$ which correlate strongly with the the logical noise strength $\overline \cN_\ell$. Of course, $\overline \cN_\ell$ is itself a function of $\cE_0$, but it is very difficult to compute even in an oversimplified model as explained in \sec{NS} (see also \sec{numerics}).

\begin{figure}
\includegraphics[scale=0.14]{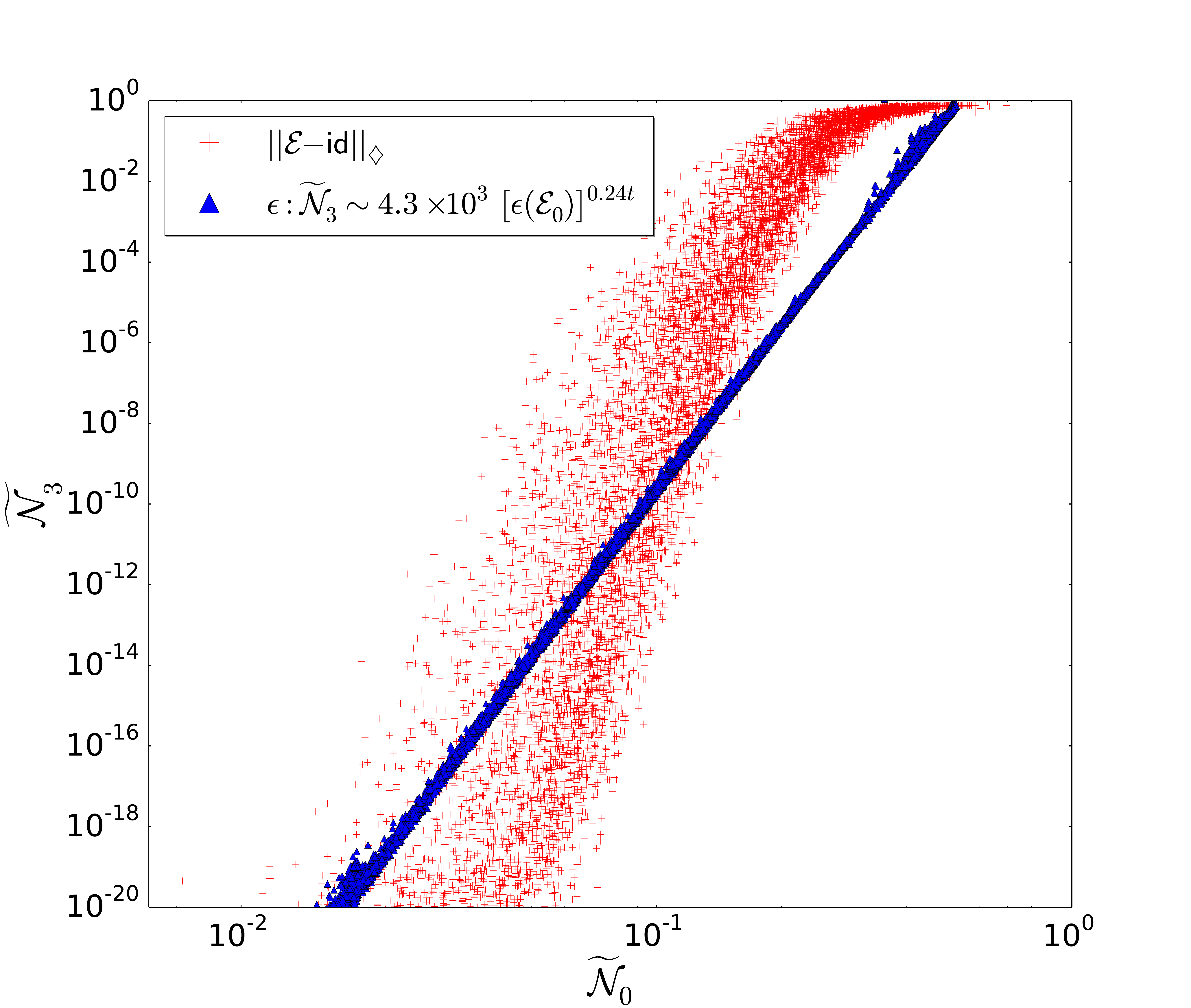}
\caption{A function $\epsilon(\cE_0)$ was computed to fit the ansatz of Eq. \ref{eq:ansatz} by minimizing the quantity in Eq. \ref{eq:least_sq_fit} over a training set of $12\times 10^4$ channels, for $\ell=1,2$ and 3 levels of concatenations. Here, we show the correlation $\epsilon(\cE_0)$ to the logical failure rate for $\ell=3$. We see that the ansatz fitted function correlates more tightly with the logical error rate compared to the diamond norm distance, shown for reference.}
\label{fig:ansatz}
\end{figure}

Motivated by the fault-tolerance accuracy threshold theorem, we make the following ansatz for the behaviour of the logical noise. For a physical noise model $\cE_0$ and given a fault-tolerant protocol family with increasing minimal distance $d$ (for the concatenated Steane code, $d = 3^\ell$), the logical noise strength decreases as
\begin{equation}
\overline \cN_d = C_\ell [\epsilon(\cE_0)]^{\alpha t} \label{eq:ansatz}
\end{equation}
where $t = \lfloor(d-1)/2\rfloor + 1, C_d$ and $\alpha$ are positive constants which are specific to the fault-tolerant scheme, while $\epsilon(\cE_0)$ is a critical parameter of the physical noise model. Our goal is thus to find this function $\epsilon$. This proceeds in two steps. We consider two arbitrary set of randomly generated physical channels, one called the \emph{training set} which in our case is the same as the one studied in \fig{scatter}. The other is called the \emph{testing set} which in our case is a different ensemble that is half the size of the training set. On the training set, we perform a least square fit of the ansatz in Eq. \ref{eq:ansatz} which minimizes the function
\begin{gather}
\sum_{\cE_0,d}\left(\log_{10}\widetilde{\cN}_{\ell} - \log_{10}C_{d} - \alpha t\log_{10}\epsilon(\cE_{0})\right)^{2} \label{eq:least_sq_fit}
\end{gather}
over the constants $\log_{10}C_d$, $\alpha$ and the $\log_{10}\epsilon$ of each channel to best fit the data. Figure \ref{fig:ansatz} shows the result of this fit for level $\ell=3$. 

\begin{figure}
\includegraphics[scale=0.355]{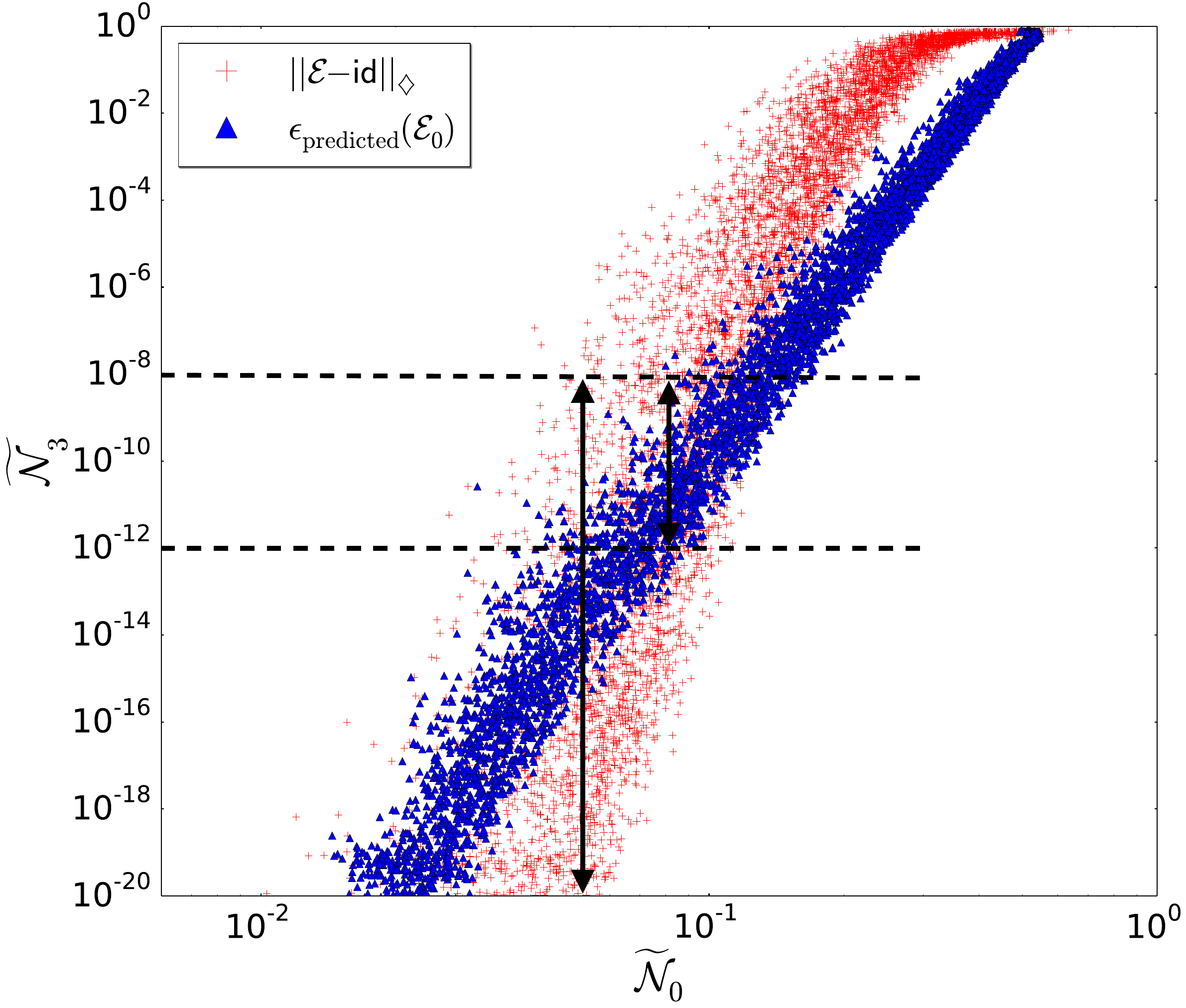}
\caption{We have trained a fully connected neural network of 100 nodes and 4 hidden layers with a rectifier (ReLU) \cite{LBH15}, to relate the numerically fitted function $\epsilon(\cE_0)$ shown in Fig. \ref{fig:ansatz}, to the parameters of the  respective the physical CPTP map $\cE_0$ in the training set. To test the efficacy of the trained neural network, we evaluated it on an entirely new ensemble of $6\times 10^4$ channels. Here, we show the logical failure rate as a function of the machine learned function $\epsilon_{\rm predicted}(\cE_0)$ and compare it to the diamond norm distance for reference. We see that the machine-learned function is a more accurate predictor of the logical error than the diamond norm distance.}
\label{fig:prediction}
\end{figure}

Then, we use one of several machine learning techniques such as \emph{kernel regression}, \emph{k-nearest neighbours} (k-NN) \cite{A92} and \emph{multi layer perceptron (MLP) regression} \cite{OM03} to relate $\epsilon(\cE_0)$ to the parameters of $\cE_0$, for all channels in the training set. The trained machine is then used to compute an estimate of $\epsilon(\cE_0)$ for channels in the testing set.  Figure \ref{fig:prediction} shows the logical noise strength as a function of this machine learned critical parameter, denoted by $\epsilon_{\rm predicted}(\cE_0)$, for noise models in the testing set. Here, the learning was done by a MLP regressor that used a $L2-$regularized square loss function and was implemented using the \emph{scikit-learn} package \cite{scikit} in Python. The machine learned parameters clearly have a better predictive power than the diamond norm distance, as shown in \fig{prediction}. For instance, the diamond norm required to achieve a logical noise rate below $10^{-8}$ can sometime yield a logical noise rate as low as $10^{-20}$. In contrast, the condition to achieve a logical noise rate $10^{-8}$ according to the machine-learned parameter also restricts the logical noise to be above $10^{-12}$. While this is a very significant improvement, it remains too coarse to be of practical interest. Note moreover that this advantage is much less pronounced when compared to the prediction obtained from infidelity (not shown).

\subsection{Discussion}

In Sec. \ref{sec:StandardMetrics} we saw that standard noise metrics can only very crudely predict the logical noise strength, while \sec{ML} further extends this conclusion to a set of optimized parameters. This shows that predicting the logical fault-rate of a fault-tolerant scheme for a given channel depends on multiple parameters of the channel $\cE_0$. We can conclude from the data presented in this section that the information about the noise cannot be compressed to a single critical parameter: the response of a fault-tolerance scheme depends critically on many parameters of the noise model. One future generalization of our approach would be to compress the information about $\cE_0$ to a few critical parameters rather than a single one. But as we begin to consider more realistic noise models with an exponentially growing number of noise parameters, our numerical experiments lead us to severely doubt that a few critical parameter will suffice to obtain an accurate predictor. 

Notwithstanding the problem of experimentally determining the noise model, we could use  numerical simulations, as we did here, to predict the logical fault rate, but as we will explain in the next section, this is generically computationally hard except in oversimplified models as used here. 

\section{Difficulty of numerical simulations}
\label{sec:numerics}

Numerical simulations have played a central role in our development and optimization of quantum error correcting schemes.  A quantum code is specified by stabilizers $S_j$: a valid code state is one for which $S_j\ket\psi = +\ket\psi$ for all $j$. In the presence of noise, the measurement of the stabilizers can yield outcomes which differ from +1. The collection of stabilizer measurement outcomes is called the syndrome, and a syndrome which is not all $+1$ signals the presence of errors. We conventionally denote the syndrome $s\in \{0,1\}$ instead of $s'\in \{+,-\}$ with the mapping $s' = (-1)^s$.  Decoding is a classical computational procedure which, given a syndrome $s$, determines the optimal recovery procedure to return the system to its initial state. The recovery is usually chosen amongst Pauli matrices, but generalizations are possible \cite{CWBL17}. 

Decoding is generically a hard problem. In the classical setting, it is well known that optimally decoding a linear code is in NP-complete \cite{BMT78}, and in the quantum setting, we have shown \cite{IP15} that the equivalent problem is in \#P-complete. This in effect means that decoding must often resort to heuristic, suboptimal methods, see, e.g., \cite{WHP03,BSV14,DP10,P08}. The decoding algorithm for concatenated codes described in \sec{NS} is a rare exception where an optimal, efficient decoding can be realized \cite{P06,F08}. In the context of fault-tolerant quantum computation, it is clear that a fast decoding algorithm is required since it has to be executed in real time \cite{CTV17}, so only efficient decoding algorithms are of interest.

The upshot is that, while optimal decoding algorithms can be numerically intractable, the decoding problem is not a bottleneck in numerical simulations since the decoding has to be efficient for any practical scheme. In other words, the goal of the numerical simulations is to study the behaviour of a noise model in a complete fault-tolerant scheme -- including its potentially sub-optimal decoding algorithm. We do not really care to know if a logical fault results from a code failure or a decoding failure. Thus, no matter what practical fault-tolerant protocol we simulate, it will have a fast decoding algorithm.

\subsection{Simulating quantum mechanics}

The two difficult parts of a numerical simulation are 1) sampling a syndrome, and 2) determining the effect of the error-correction procedure on the logical qubit. These are inherently quantum mechanical problems and have no classical counterpart. Let us indeed consider the classical setting first (we will describe syndrome-based decoding). 

In a numerical simulation of classical error correction, we prepare a codeword and simulate its noisy transmission. Given a received noisy bit string, the syndrome consists of parities of subsets of the received bits, which can be computed efficiently. The decoder then takes as input this syndrome and outputs the optimal recovery, i.e. the optimal sequence of bits to flip in order to recover the initial codeword. We can then check if this decoded codeword coincides with the initial codeword, which had been kept in memory for the sake of the simulation. Repeating this procedure enables us to estimate the fault rate.

It comes out of the above description that the syndrome does not need to be sampled: instead, it is the error itself which is sampled. In other words, we directly simulate the error process of, e.g., flipping each transmitted bit with some probability $p$. The syndrome is a function of the resulting noisy bit string, there is no additional randomness involved in producing it. What also comes out of the above description is that each run of the algorithm will either result in a failure or a success, and that determining which occurred is computationally trivial. 

In the quantum setting, it is generically not possible to sample  the error because the noise model isn't always stochastic in nature. A simple example is the systematic rotation channel, where each qubit undergoes a small rotation $U_\theta = e^{i\frac{\theta}{2} X} = \cos\frac \theta 2 I + i\sin\frac\theta 2 X$. We can think of this error as a {\em coherent superposition} of having no error $I$ {\em and} having a bit flip error $X$. This is distinct from a stochastic model having no error $I$ {\em or} having a bit flip error $X$. Under the coherent error model, the syndrome has an undetermined value and we are forced to numerically simulate its measurement. 

To illustrate this, consider a  3-qubit code with stabilizers $ZZI$ and $IZZ$, and with corresponding logical states $\ket{\bar 0} = \ket{000}$ and $\ket{\bar 1} = \ket{111}$. Starting in an arbitrary initial code state $\ket{\bar\psi} = \alpha\ket{\bar 0} + \beta\ket{\bar 1}$, the error model will result in the state $U_\theta^{\otimes 3}\ket{\bar\psi}$. Upon measurement, the syndromes have the following probabilities
\begin{align}
\Pr(s = 00) &= (\cos\tfrac\theta 2)^6 + (\sin\tfrac\theta 2)^6,\\
\Pr(s = 01) &= \Pr(s = 10) = \Pr(s = 11) \\
&= (\cos\tfrac\theta 2)^4(\sin\tfrac\theta 2)^2 + (\cos\tfrac\theta 2)^2(\sin\tfrac\theta 2)^4. \nonumber
\end{align}
After error correction, the  syndrome ++ will result in the state
\begin{equation}
\ket{\bar\psi^{s = 00}} \propto \left[(\cos\tfrac\theta 2)^3 I  -i (\sin\tfrac\theta 2)^3 ZZZ\right]\ket{\bar\psi}
\end{equation}
while the other three syndromes would produce the state
\begin{equation}
\ket{\bar\psi^{s}} \propto \left[(\cos\tfrac\theta 2)^2(\sin\tfrac\theta 2) I  +i (\sin\tfrac\theta 2)^2(\cos\tfrac\theta 2) ZZZ\right]\ket{\bar\psi}.
\end{equation}
Thus, we see that the syndrome value is not determined by the error, so it must be sampled, and that in all cases the final state is not exactly equal to the original state, nor is it orthogonal -- a residual logical error $\cE_1^s$ remains. In this example, the probabilities and residual logical error could be computed analytically, but in general this will not be possible. For most codes and under generic single qubit noise models $\cE_0$, simulating the syndrome measurement and evaluating the resulting logical error $\cE_1^s$ can only be done by simulating an $n$-qubit density matrix, with memory requirement $4^n$. The algorithm presented in \sec{NS} uses special structure of concatenated codes to circumvent this exponential cost, and the algorithm of \cite{DP17} uses the tensor-network structure of the surface code to achieve complexity $8^{\sqrt n}$. It is not clear at all whether these simulations can be realized using a memory of size less than $4^n$  when we include more realistic noise models where gates and measurements are also noisy. 

There exist a class of quantum channels with a stochastic interpretation, for which numerical simulations become essentially identical to the classical case. These are Pauli noise model, and have been used in the overwhelming majority of numerical simulations to date. A Pauli channel $\cP$ maps a density matrix $\rho$ to $\cP(\rho) = \sum_P p_P P\rho P$, where the sum runs over all the Hermitian (multi-qubit) Pauli operators $P$, and the $p_P$ are non-negative and sum to 1, i.e. they form a probability distribution. In other words, Pauli channels are CPTP maps whose $\chi$ matrix is diagonal.  In a complete Pauli noise model, every component $\cG$ of a quantum circuit (preparation, gate, or measurement) is modelled by the ideal component, followed (or preceded for a measurement) by a Pauli channel $\cP_\cG$. A Pauli noise model is thus a stochastic noise model. Indeed, we can give it the interpretation that every time a  gate $\cG$ should be applied in the ideal circuit, there is a probability $p_{P|\cG}$ that gate $P\cG$ is applied instead. 

Because the commutation of Pauli matrices follow a simple pattern, it is easy to determine the syndrome given a sampled Pauli error. Likewise, the combination of the error and the correction will either result in the logical identity or a non-trivial logical gate, which can easily be determined. This is a simple consequence of the Gottesman-Knill theorem. Thus, for the sake of numerical simulations, we see that Pauli noise models behave essentially like classical channels.  Unfortunately, the noise produced in most hardware cannot be well approximated by Pauli noise. A common strategy is to use a Pauli noise model as a proxy to the device's noise only for the sake of numerical simulations. But unfortunately, this yields very inaccurate predictions of the logical fault rate \cite{DP17}. Thus, while numerical simulations using Pauli noise are efficient and can provide a coarse characterization of a fault-tolerant scheme, they cannot be used to predict its response to a physically realistic noise model.

\subsection{Importance of outliers}

In addition to the difficulties of simulating quantum systems described above, numerical simulations of classical and quantum error correction face the inherent difficulty of characterizing rare events. Let us begin by estimating the logical error rate that we need to characterize. According to \cite{AGP06}, it takes $\sim 34^k$ gates to implement one level-$k$ logical gate. Assuming the typical MHz clock cycle of solid state qubits and two levels of concatenation results in a 1kHz logical  gate rate, so the logical circuit can reach a depth of nearly one billion in one day. Gates (including identity) are applied in parallel, so for a 1000 logical qubit device, we get $10^{12}$ gates per day. So if our goal is to protect a one-day quantum computation, we need to characterize the logical noise down to accuracy $10^{-12}$ assuming that it builds up linearly.\footnote{For incoherent noise, two folk results appear to contradict each other here. On the one hand, it is often said that stochastic errors build up like a random walk, so that in the current example, a logical fault rate of $10^{-6}$ would suffice. On the other hand, there is a widespread belief that after error correction, the logical channel is Pauli. But clearly, a single logical Pauli error is enough to invalidate the whole computation, so we again require a $10^{-12}$ target.}

Estimating such a small number reliably is not a simple task. This is particularly true when the logical failure rate is dominated by atypical syndromes, i.e., outliers. To understand this, consider two extreme types of syndromes for a minimum-distance $d$ code used on a stochastic channel in the low error regime $p\ll 1$. On the one hand, the trivial syndrome occurs with probability $\Pr(s=0) \simeq (1-p)^n \in \cO(1)$. The optimal recovery in this case is the identity, and the next most-likely error is a logical operator, whose probability is $\cO(p^d)$. Thus, the residual logical error when the trivial syndrome is observed is $\cN(\cE^{s=0}) \in \cO(p^d)$. On the other hand, consider a syndrome $s^*$ which signals the presence of an error $E$ of weight roughly $d/2$. Such a syndrome has a much lower probability $\Pr(s^*) \in \cO(p^{\frac{d}{2}})$. But in that case, there exist another inequivalent error $E'$ of weight roughly $d/2$ that is compatible with the syndrome. This happens when the combination of the two errors $E$ and $E'$ form a logical operator. So in this case, the probability of misdiagnosing the error is $\cO(1)$ because the two inequivalent alternative are roughly equiprobable. So the residual logical error in the event of such an unlikely syndrome is $\cN(\cE^{s^*}) \in\cO(1)$. Taking the contributions from the two types of syndromes to the total average logical error yields
\begin{align}
\overline \cN &= \Pr(s=0) \cN(\cE^{s=0}) +\Pr(s^*) \cN(\cE^{s^*}) \\
& \in \cO(p^d+p^{\frac d2}) = \cO(d^{\frac d2}).
\end{align}
We see that the average logical noise strength is totally dominated by syndromes which occur with a much lower probability -- the outliers.

\begin{figure}
\begin{center}
\includegraphics[scale=0.4]{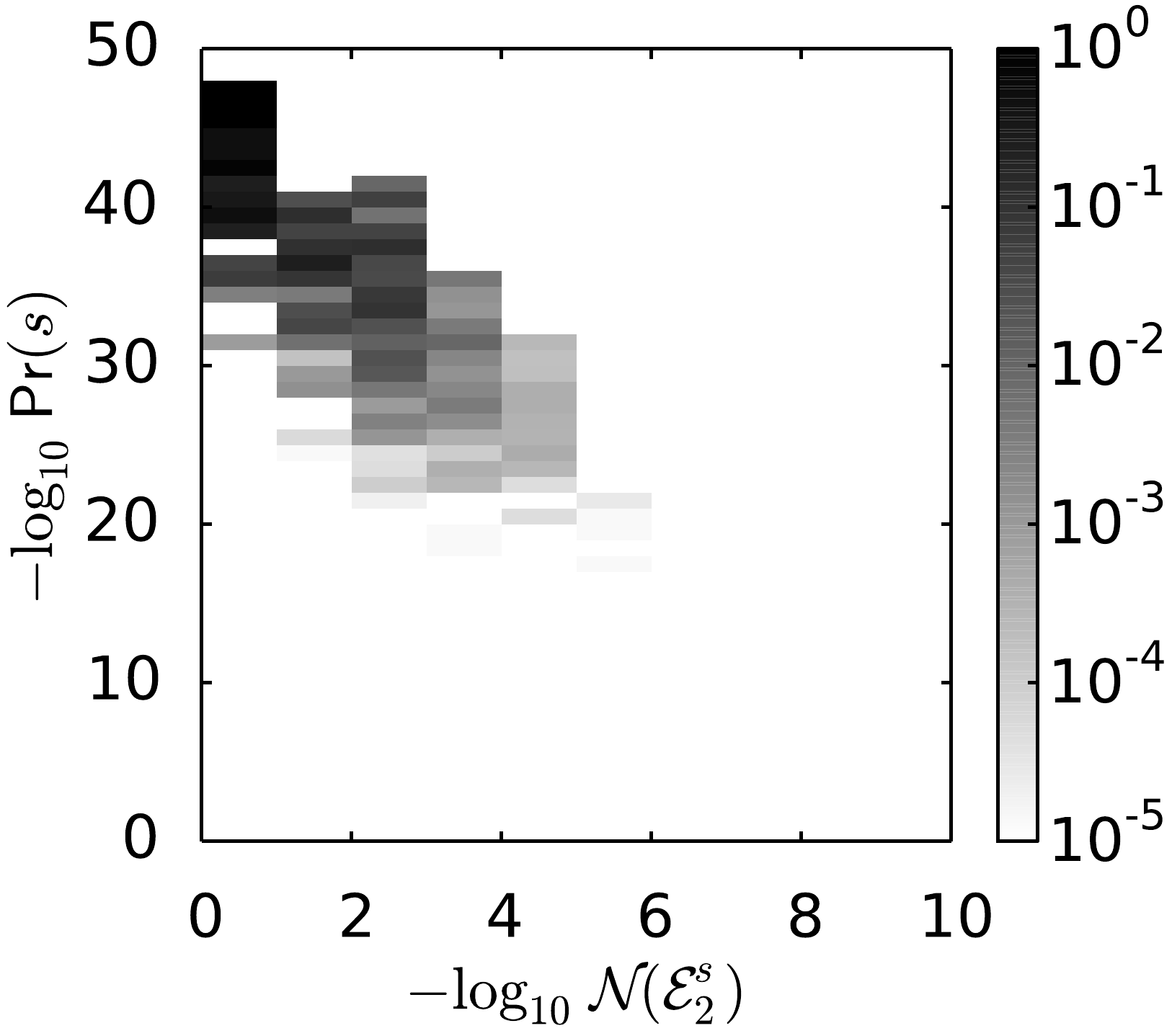}
\caption{Density plot showing the fraction of syndromes with a given probability $\Pr(s)$ and resulting in given logical noise strength $\cN(\cE^s_k)$. These syndromes are measured for a level 2 concatenated Steane code under a randomly generated physical noise process as described in \sec{NS}, with $\delta = 0.02$. The density in the plot is proportional to $C(\cT)$ in Eq. \ref{eq:combinatorial}. The majority of is syndromes result in a high ($\sim 1-0.01$) logical noise strength, but they cannot be observed in Monte Carlo simulations with reasonable sample size ($N\sim10^6-10^{10}$)  because their probability is too low ($\lesssim 10^{-20}$).}
\label{fig:outliers}
\end{center}
\end{figure}

What the above analysis neglects are combinatorial factors indicating how many errors of each type exist. As in the above analysis, suppose we organize the syndromes into different types $\cT$, with each syndrome $s$ of a given type $\cT$ having similar probability of occurring $\Pr(s) = \Pr_{\cT}$ and result in the same  residual logical noise strength $\cN(\cE^s) = \cN_{\cT}$. The exact expression for the average logical noise strength is
\begin{equation}
\sum_{\cT\in {\rm types}} C(\cT){\rm Pr}_{\cT}\cN_{\cT}. \label{eq:combinatorial}
\end{equation}
where $C(\cT)$ denotes the number of errors of a given type, and is related to the weight enumerator of the code. Fig. \ref{fig:outliers} shows the (normalized) combinatorial factor $C(\cT)$. There, we clearly see that the overwhelming majority of syndromes lead to a high logical fault rate, but on the other hand they have an exceedingly low probability of occurring. These constitute the outliers described in the above paragraph, and their presence is observed in our numerical simulations. In particular, we have observed that Monte Carlo simulations using a small number $N$ of samples tends to underestimate the logical failure rate. The estimated failure rate $\widetilde \cN = \frac{1}{N} \sum_{j=1}^N \cN(\cE^{s_j})$ tends to make sudden positive jumps as a function of $N$, see \fig{importance_sampling} a). This can be easily explained by the existence of outliers:  the sample underestimates the logical fault rate until an outlier is sampled, which occurs very infrequently. 

So formally, the results shown on \fig{scatter} cannot be trusted below $\widetilde \cN \leq 10^{-4}$ because the Monte Carlo sample size was only $10^4$ -- the true fault rate could be much larger but we simply haven't sampled long enough to catch the outliers. To assess with high confidence that a fault tolerant scheme produces a logical failure rate $10^{-12}$ for a given noise model, one should in principle collect $10^{12}$ Monte Carlo samples. Note that our goal in \fig{scatter} was not to get a precise estimate for any given channel, but instead  grasp how differently distinct channels behave. The fact that the depolarization and rotation channels show statistical fluctuations which are much less than the difference between them makes us confident that our conclusions regarding the variation of the logical fault rate for different physical channels are essentially correct. 

\subsection{Importance sampling}
\label{sec:IS}

Importance sampling \cite{LMT09} was developed to speed-up the sampling of rare events. Abstractly, consider a random variable $X$ taking values $x_j$ with probability $\Pr(j)$, and assume without loss of generality that $\Pr(j)>0$. For an arbitrary probability distribution $q_j$, define another random variable $Y$ taking values $y_j = x_j \Pr(j)/q_i$ with probability $q_j$. Clearly, $X$ and $Y$ have the same average. So in particular we can estimate $\langle X\rangle$ by sampling $Y$. By suitably choosing the probability $q_j$, the random variable $Y$ can have a smaller variance than $X$, so sampling $Y$ would converge faster. For a positive random variable $X$, a trivial example illustrating this is setting $q_j = x_j/\langle X\rangle$, in which case a single sample of $Y$ yields the expectation value of $X$. This example is not realistic of course because it requires knowledge of the quantity $\langle X\rangle$ we seek to estimate. 

In the setting of classical error correction, importance sampling can be used by increasing the probability of the outliers. Of course we do not know ahead of time what the outliers are, but several techniques can be adopted to produce the desired effect. These techniques are directly applicable to quantum error correction with Pauli noise models \cite{BV13,LGD17,TLGD17}, where we can reassign probabilities to the various Pauli errors. 

But for non-stochastic noise models, importance sampling is not straightforward because there is no probability associated to errors. But there are probabilities associated to syndromes, so we can modify those to realize importance sampling. In other words, the syndromes will be picked not according to Born's rule of quantum mechanics $\Pr(s)$, but using a different probability distribution $Q(s)$. We shall refer to $Q$ as the \emph{importance distribution} and the corresponding sampling algorithm as the \emph{importance sampler}. Likewise $\Pr(s)$ is referred to as the \emph{true distribution} and the corresponding sampling algorithm as the \emph{direct sampler}. 

\begin{figure*}
\begin{center}
\subfloat[]{\includegraphics[scale=0.14]{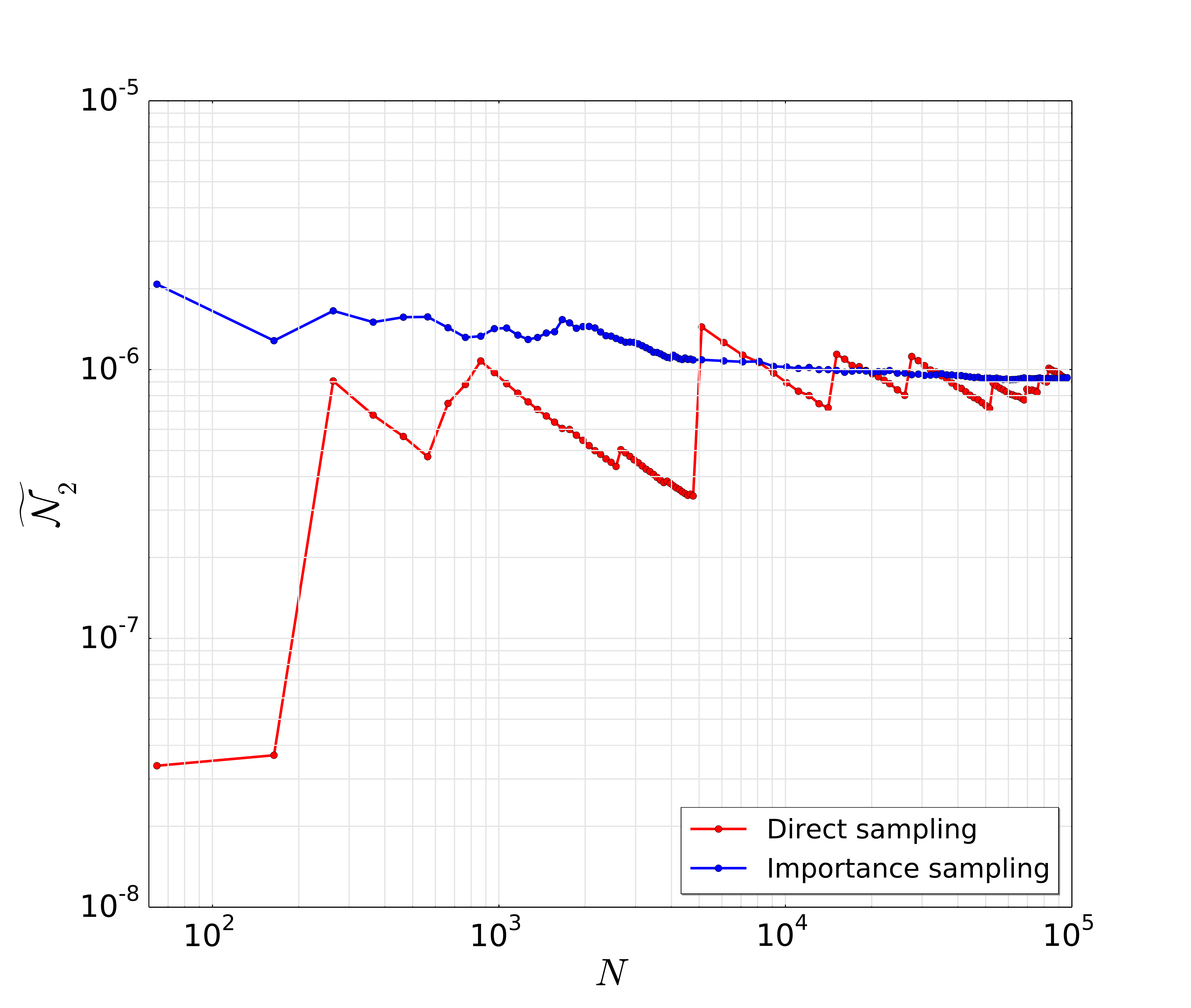}}
\subfloat[]{\includegraphics[scale=0.14]{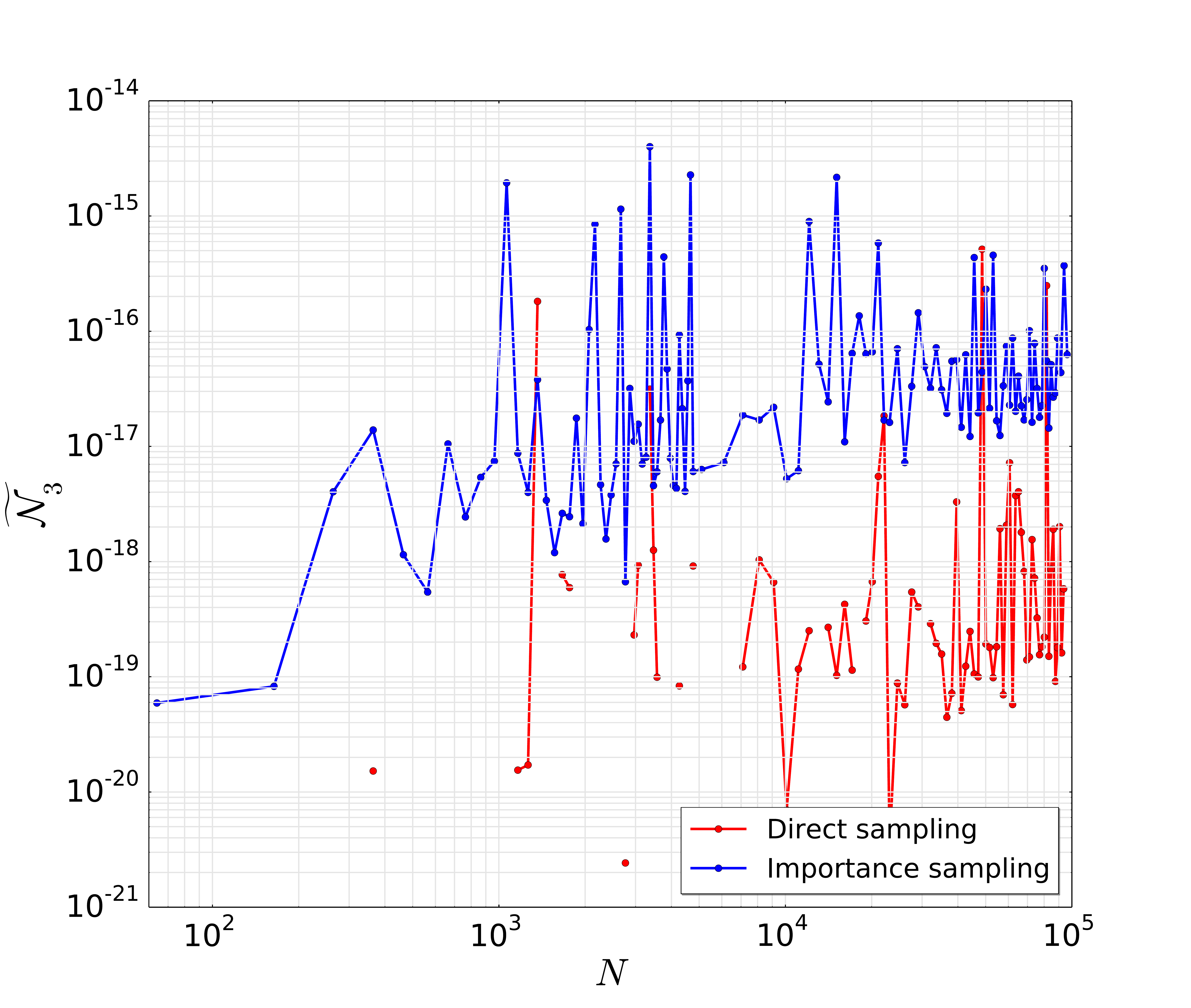}}
\caption{Average logical error as estimated by direct sampling (red) and importance sampling (blue) as a function of the sample size for a random (fixed) physical channel $\cE_0$. In a) the logical error rate is calculated for a $49$ qubit (level 2) concatenated Steane code, while in b) it is for a $343$ qubit (level 3) concatenated Steane code. The direct sampler underestimates the logical error rate with small samples, and makes sudden positive jumps when an outlier is sampled. The importance sampler favors outliers and thus converges to the right value using a smaller sample in a). The advantage of importance sampling is less obvious in b).}
\label{fig:importance_sampling}
\end{center}
\end{figure*}

Since our goal is to increase the probability of the outliers, we choose a distribution which limits the probability of the trivial syndrome in favor of the other syndromes. For instance, we can set
\begin{gather}
Q(s) = \dfrac{\Pr(s)^{\beta}}{Z} \label{eq:imp_dist_power_law}
\end{gather}
for some power $0< \beta \leq 1$ and some normalization factor $Z$, where $\beta$ is chosen such that
\begin{gather}
Q(0) = \text{min}\left(\Pr(0), \frac 12 \right). \label{eq:imp_dist_cutoff}
\end{gather}
Figure \ref{fig:importance_sampling} compares the estimated average obtained by a direct sampler and an importance sampler as a function of the sample size, for a level$-2$ as well as level$-3$ concatenated Steane code under a randomly generated physical noise process. In \fig{importance_sampling} a), the estimate of the direct sampler is strongly affected by the encounter of outlier syndromes as can be seen in the sudden positive jumps in the estimated logical fault rate. On the other hand, the importance sampler converges to the true average, i.e, the same as the direct sampler for large sample sizes, even at relatively small sample sizes. For that specific example, an importance sample of size $N\sim 5\times 10^3$ yields the same statistical fluctuation as a direct sample of size $N\sim 10^5$. 

While this is a significant improvement, we cannot conclude that the importance distribution we have chosen always provides an advantage. For instance, \fig{importance_sampling} b) uses the same importance distribution on the same channel to estimate the average logical error for level $\ell = 3$ but results in a much less convincing advantage. And unfortunately, the only way we can tell for sure that an importance sampler converges more rapidly to the true average is to produce a much larger direct sample to compare with. Thus, at this stage, importance sampling of quantum error correction consist more of an art than a science.


\subsection{Discussion}

Despite using an oversimplified noise model, the numerical simulations performed for this article required 40 milliseconds per round for two concatenation layers of Steane's code.  This is roughly 40 times slower than the anticipated time required by the hardware to perform one error-correction round. While this difference can easily be compensated by performing simulations in parallel, the simulation of a full noise model -- with noisy gates and measurements and non-Pauli errors -- will require far more resources. A recent record shattering experiment used a supercomputer for two days in order to simulate a 56-qubit circuit of depth 23, using up to 3 TB of memory \cite{PGNH17}. This circuit is smaller than the one required by two concatenation layers of Steane's  code. Moreover, it uses only pure states, so in terms of memory and number of operations it is closer to a 23-qubit mixed state simulation. 

Just like the surface code simulation \cite{DP17}, this 56-qubit simulation used tensor networks to achieve a computational speed-up, and surely other such tricks will be developed in the future. But unless a numerical revolution occurs, it seems inconceivable that classical simulations could be used to verify with confidence that a given fault-tolerant scheme achieves the targeted logical fault rate $10^{-12}$ required to reliably run a modest-size quantum computer for a day. But, by definition, this task could be accomplished in one day on a modest quantum computer.

\section{Discussion and Outlook}

Building a quantum computer capable of outperforming classical supercomputers will require further developing and optimizing fault tolerant protocols. While simple  optimizations can be assessed by numerical simulations, we have argued in this article that reaching the level of accuracy of interest to optimize a protocol for a modest quantum computer is far beyond the reach of numerical simulations. The reasons we invoked are
\begin{enumerate}
\item The difficulty of characterizing the noise in hardware;
\item The high sensitivity of fault-tolerant protocols to the parameters of the noise model; and
\item The difficulty of numerically simulating fault-tolerant protocols.
\end{enumerate}

On the other hand, all of these difficulties disappear if we directly assess the quality of a fault-tolerant protocol on a quantum computer. Concretely, this could be realized by elevating the protocols used to characterize the noise strength of physical qubits to characterizing the noise strength of logical qubits. For instance, we could perform logical tomography \cite{ZLS12,G97}, or logical randomized benchmarking \cite{CGFF17, CMBS16}, or logical gate set tomography \cite{KGK17,KLR08,G15}, etc. The feasibility of these protocols follows from the fact that we are only interested in characterizing the validity of the gates to the extent that we are going to use them. If, as in the examples above, our goal is to secure a one-day quantum computation to some constant success probability, that a few days of logical characterization are sufficient to achieve it.  

While it will certainly not replace the need for numerical simulations and experimental noise characterization, we believe that the direct experimental characterization of fault-tolerant scheme advocated here will at least be one important ingredient in the fault-tolerant optimization toolkit. Experimental noise characterization has been critical for reducing errors in physical devices because it provides insight about its physical origin, and there is no doubt that this will continue play an important role. But fault-tolerant protocols are not concerned with reducing errors in the hardware, their purpose is to cope with errors at the software level, so do not benefit from a physical understanding of the noise mechanism. 

Likewise, numerical simulations have been critical for developing new fault-tolerant protocols and obtaining crude assessment of their performance. There is no doubt that numerical simulation will continue to provide guidance into the theory of fault tolerance, but compared to actual experiences they will be of very little use for the purpose of optimizing a protocol to a given hardware. Numerical simulations have been extensively used to estimate the logical fault rate $\overline \cN$ as a function of a physical noise parameters $p$ of a simple noise model. This has little bearing on the problem of estimating the logical fault rate for a realistic noise models encompassing numerous fixed parameters. In particular, the protocol with the best scaling as a function of $p$ is not necessarily the optimal protocol for some set of fixed noise parameters and for a fixed target logical fault rate. 

Perhaps the most powerful optimization tools will use a classical-quantum hybrid, where the quantum computer is used as a sub-routine to the classical simulation. In fact, as we were just finalizing this article, similar ideas were proposed in a preprint \cite{JROC17} where a quantum computer is used as a subroutine in a classical optimization procedure to numerically optimize a fault tolerant protocol to a noisy device. The general task of working out a concrete optimization toolchain is a challenging problem which is left open for future research, as the needs develop.

\section{Acknowledgements} 
We thank Marcus da Silva, Steve Flammia, Robin Blume-Kohout and Stephen Bartlett for raising concerns during the evolution of this project. This work was supported by the Army Research Office contract number W911NF-14-C-0048. 

\bibliography{refs}

\appendix

\section{Definitions for natural error metrics} \label{app:naturalMetrics}
In this section, we will specify the definitions of the natural error metrics which we have mentioned in Sec. \ref{sec:StandardMetrics} to quantify the strength of noise in quantum channels. In this section, we will use $\cE$ to denote a single qubit CPTP map and $\cJ$ to denote its Choi-matrix (see Sec. \ref{app:numerics}). $\cE_{\rm id}$ denotes a trivial CPTP map, that maps any quantum state to itself. Additionally, the representation of $\cE$ as a $4\times 4$ matrix $\Gamma$ with real non-negative entries, specified by
\begin{gather}
\Gamma = \dfrac{1}{2}\sum_{i,j = 1}^{4}\Tr(\cE(P_{i})P_{j})|i\rangle\langle j| \label{eq:pauli_liouville}
\end{gather}
is called the Pauli-Liouville representation of $\cE$ \cite{KSRJ14}. In the above expression, $i$ and $j$ label the different Pauli matrices, $P_{i}, P_{j} \in \{I, X, Y, Z\}$ whereas $|i\rangle, |j\rangle$ are computational basis states.
\begin{enumerate}
\item Diamond norm distance: By this we refer to the Diamond norm distance between $\cE$ and $\cE_{\rm id}$ denoted by $||\cE - \cE_{\rm id}||_{\diamondsuit}$, defined as
\begin{gather}
||\cE - \cE_{\rm id}||_{\diamondsuit} = \dfrac{1}{2}\mathsf{sup}_{\rho}||\widetilde{\cE}(\rho) - \rho||_{1} \label{eq:dnorm_def},
\end{gather}
where $\widetilde{\cE}$ is an extension of the channel to multi qubit states in such a way that the channel only acts trivially on all but the first qubit, on which its action is given by $\cE$. We use the semi-definite program in \cite{W09} to compute $||\cE - \cE_{\rm id}||_{\diamondsuit}$.
\item Infidelity: By this we refer to the infidelity of $\cJ$ to the bell state, denoted by $1 - \cF$ where
\begin{flalign}
\cF &= \dfrac{1}{2}\left(\langle00| + \langle11|)\cJ(|00\rangle + |11\rangle\right) \nonumber \\
&= 1 - \dfrac{1}{2}\left(\cJ_{1,1} + \cJ_{1,4} + \cJ_{4,1} + \cJ_{3,3}\right) \label{eq:infidelity_choi} \\
&= \dfrac{1}{4}\Tr(\Gamma) \label{eq:infidelity_process}.
\end{flalign}
$1-\cF$ is popularly referred to as the \emph{entanglement infidelity} \cite{B96,N96} and it differs from the average infidelity of \cite{KSRJ14,CWBL17} by a constant factor.
\item $L1, L2-$norms: We have used $\cJ$ to define the $L1, L2$ norms for $\cE$ \cite{G05}. The $L1-$norm or \emph{Trace norm} of $\cE$ is specified by $||\cJ - \cJ_{\rm id}||_{1} = \Tr(\sqrt{(\cJ - \cJ_{\rm id})^{\dagger}(\cJ - \cJ_{\rm id})})$. Likewise, we refer to the $L2-$norm or \emph{Frobenius norm} of $\cE$ to be specified by $||\cJ - \cJ_{\rm id}||_{2} = \sqrt{\Tr((\cJ - \cJ_{\rm id})^{\dagger}(\cJ - \cJ_{\rm id}))}.$
\item Worst case error: The worst case error for $\cE$, denoted by $p_{\rm err}$ is defined \cite{AGP07} to be the solution of the following optimization problem.
\begin{align}
& {\rm max} & x \label{eq:worst_case_error} \\
& \text{subject to:} & 0\leq x\leq 1\nonumber \\
&& \left(\cJ - (1-x)\cJ_{\rm id}\right) \succcurlyeq 0,\nonumber
\end{align}
where the last constraint indicates that $\left(\cJ - (1-p)\cJ_{\rm id}\right)$ must be a positive semidefinite matrix, i.e, have non-negative eigenvalues. For stochastic channels, $p_{err}$ is equal to the total probability of non-identity Krauss operators.
\end{enumerate}

\section{Details of numerical simulations} \label{app:numerics}
In this section, we provide details of the numerical simulations outlined in Sec. \ref{sec:NS}. Before proceeding, we provide a few definitions. Let $\bar{\rho}$ be an encoded state of the Steane code with stabilizer $\cS$. Upon the application of an i.i.d channel $\cE^{\otimes 7}_{0}$, where $\cE_{0}$ is a single qubit CPTP map, we get $\rho_{\rm noisy}$. The subscript 0 makes reference to the fact that these are physical noise process. We then apply the quantum error correction circuit to $\rho_{\rm noisy}$. Let $\Pi_0$ denote the projector on to the code space and $\Pi_s$ denote the projector on to the syndrome space of $s$, expressed as
\begin{gather}
\Pi_{s} = \prod_{j=1}^{n-k}\left(\dfrac{\II + (-1)^{s_j}S_j}{2}\right) \label{eq:syndproj}
\end{gather}
where $n = 7, k = 1$ for the Steane code and $s_j\in\{0,1\}$ is the $j$th syndrome bit. Upon expanding the above projector, we obtain
\begin{gather}
\Pi_{s} = \sum_{S \in \cS}\phi_{S} S \label{eq:sydproj_expanded},
\end{gather}
where $\phi^{s}_{S} \in \{+1, -1\}$ is the parity of syndrome bits of $s$ whose corresponding stabilizer generators appear in the decomposition of $S$. The probability of measuring a syndrome is just $\Pr (s) = \Tr(\rho_{\rm noisy}  \Pi_s)$. Let $T_s$ be a Pauli operator that takes a state from the syndrome $s$ subspace to the code space, i.e, $\Pi_s = T_s   \Pi_0   T_s$, in other words, $\rho_{\rm noisy}^{s} = T_s   \Pi_s   \rho_{\rm noisy}   \Pi_s   T_s$ lies in the code space. In order to obtain the correct logical state with high probability, we must apply a logical Pauli operator $\bar{Q}$ that maximizes the fidelity
\begin{gather}
F(\bar{\rho}, \rho_{\rm noisy}^{s, Q}) = \dfrac{\Tr(\bar{\rho}   \rho_{\rm noisy}^{s, Q})}{\Pr (s)} \label{eq:qecc_fidelity}
\end{gather}
where $\rho_{\rm noisy}^{s, Q} = \bar{Q}\rho_{\rm noisy}^{s} \bar{Q}$. For stochastic noise models such as the depolarizing channel, the above described quantum error correction scheme is optimal and known as \emph{maximum likelihood decoding}. Finally, the optimal correction $\bar{Q}_{\rm max}$ is applied and the output of the quantum error correcting circuit can be mapped to a single qubit state $\rho^{\prime}$ given by
\begin{gather}
\rho^{\prime} = \sum_{P \in \{I, X, Y, Z\}}\Tr(\Pi_{0}  \bar{P}  \rho_{\rm noisy}^{s, Q_{\rm max}})\thickspace P \label{eq:qecc_unencoded}.
\end{gather}
Hence the combined effect of encoding map, noise process and quantum error correction can be encapsulated in a single qubit effective channel \cite{RDM02} denoted by $\cE^{s}_{1}$ whose action is given by $\cE^{s}_{1}:\rho\mapsto\rho^{\prime}$.

In order to extract the description of the effective logical channel, we make use of another tool, an isomorphism between channels and states, called the Choi-Jamio\l kowski isomorphism \cite{J72,C75}. Under this isomorphism, a single qubit channel $\cE$ is expressed using the two-qubit state $\cJ(\cE)$, also called the \emph{Choi matrix} of $\cE$, given by $\cJ(\cE) = \dfrac{1}{4}\sum_{i = 1}^{4}\cE(P_i) \otimes P_{i}^T$, where $P_{i}$ are Pauli matrices. Furthermore, when $\cE$ is represented as a Pauli-Liouville matrix $\Gamma$ as in Eq. \ref{eq:pauli_liouville}, where $\Gamma_{ij} = \Tr(\cE(P_i)  P_j)$, we have
\begin{gather}
\Gamma_{ij} = \Tr(\cJ(\cE)  (P_j\otimes P^T_i)) \label{eq:choi_to_process}.
\end{gather}
Our goal is to construct the two qubit state that corresponds to the Choi matrix of the effective logical channel $\cE_{1}^{s}$. Hence in the above equation, we must substitute for $\cE$, the composition of encoding map, the noise process and the quantum error correction circuit.
In our simulation, we reconstruct this composition. To start, we prepare a $8-$qubit state $\rho_{0}$ which consists of a maximally entangled state between an encoded qubit in the Steane code and a reference qubit. The reference qubit is noiseless while the qubits of the Steane code undergo the i.i.d channel $\cE^{\otimes 7}_{0}$ whose Pauli-Liouville matrix takes the representation $\Gamma_{0}^{\otimes 7}$, where $\Gamma_{0}$ is the process matrix representing $\cE_{0}$. Consequently,
\begin{gather}
\rho_{\rm noisy} = \dfrac{1}{4}\sum_{u}\cE^{\otimes 7}_{0}(\Pi_{0}   \bar{P}_{u}  \Pi_{0})\otimes P^{T}_{u}
\end{gather}
and the probability of a syndrome takes the following simple form.
\begin{flalign}
\text{Pr}(s) &= \Tr(\Pi_{0}  \Pi_{s}) \nonumber \\
&= \dfrac{1}{2^{n-k}}\sum_{\substack{i\\P_{i} \in \cS}}\sum_{\substack{j\\P_{j} \in \cS}}\left[\Gamma_{0}^{\otimes 7}\right]_{ij} \phi^{s}_{j} \label{eq:synd_prob},
\end{flalign}
where $\phi^{s}_{j}$ is a phase associated with the $j$-th stabilizer as in Eq. \ref{eq:sydproj_expanded}. Then, we simulate a syndrome measurement by numerically computing $\text{Pr}(s)$ and selecting a random syndrome according to the distribution given by $\text{Pr}( )$. Once a syndrome is chosen, we need to compute the fidelities in Eq.~\ref{eq:qecc_fidelity} to determine the optimal logical correction. It is easy to see that
\begin{gather}
F(\rho_{0}, \rho_{\rm noisy}^{s, Q}) = \dfrac{1}{\Pr (s)}\sum_{\substack{i\\P_{u} \in \cS}}\sum_{\substack{j\\P_{j} \in Q  P_{u}  Q  \cS}}\left[\Gamma_{0}^{\otimes 7}\right]_{ij} \phi^{s}_{j} \label{eq:qeccc_fid_process},
\end{gather}
where $Q  P_{u}  Q  \cS$ refers to the set of all Pauli operators obtained by multiplying every stabilizer in $\cS$ to $Q  P_{u}  Q$. Finally, the output of the quantum error correction circuit along with the reference qubit, can be mapped to the two qubit entangled state, using Eq. \ref{eq:qecc_unencoded}, that corresponds to the Choi matrix of the effective logical channel $\cE^{s}_{1}$. Using the mapping in Eq. \ref{eq:choi_to_process}, we can immediately write the Pauli-Liouville matrix $\Gamma^{s}_{1}$ corresponding to $\cE_{1}$ as
\begin{gather}
\left[\Gamma^{s}_{1}\right]_{a,b} = \dfrac{1}{2^{n-k}}\sum_{\substack{i\\P_{i} \in \bar{P}_a \cS}}\sum_{\substack{j\\P_{j} \in Q_{\rm max}  \bar{P}_b  Q_{\rm max} \cS}}\left[\Gamma^{\otimes 7}_{0}\right]_{ij} \phi^{s}_{j}. \label{eq:eff_channel}
\end{gather}
The derivation of $\Gamma^{s}_{1}$ terminates one instance of the simulation. Note that computing the quantities in Eqs. \ref{eq:synd_prob} through \ref{eq:eff_channel} require at most $4^{7}$ elementary operations. As described in Sec. \ref{sec:NS}, to obtain an effective logical channel for $\ell = 2$ levels of concatenation, we need to repeat the above simulation to calculate $\Gamma^{s}_{2}$ using Eq. \ref{eq:eff_channel} where $\Gamma_{1}^{s}$ is replaced by $\Gamma^{s}_{2}$ and each of the $\Gamma_{0}$ are replaced by an effective channel $\Gamma^{s_j}_{1}$ where $s_j$ are part of the syndrome history of $s$. The acquisition of an effective channel for  $\ell$ concatenation level, $\cE^{s}_{\ell}$, requires $7^{\ell - 1}$ error correction steps at level 1, $7^{\ell - 2}$ error correction steps at level 2, and so on. Hence, the time complexity of computing $\cE^{s}_{\ell}$ is $4^{7} 7^{\ell}$, which is linear in the number of physical qubits of the concatenated code.
\end{document}